\documentstyle[12pt]{article}
\newcommand{\be}{\begin{equation}}
\newcommand{\ee}{\end{equation}}
\newcommand{\beqa}{\begin{eqnarray}}
\newcommand{\eeqa}{\end{eqnarray}}
\begin{document}

\begin{center}
{\bf Aspects of generalized Calogero model }
\end{center}
\bigskip
\begin{center}
S.Meljanac$^{a}${\footnote{e-mail: meljanac@irb.hr}}, 
 M.Milekovi\'{c} $^{b}$ {\footnote{e-mail: marijan@phy.hr }},
    A. Samsarov$^{a}$ {\footnote{e-mail: andjelo.samsarov@irb.hr }}\\ [3mm]                                 

\bigskip

$^{a}$ Rudjer Bo\v{s}kovi\'c Institute, Bijeni\v cka  c.54, HR-10002 Zagreb,
Croatia\\[3mm]

$^{b}$ Theoretical Physics Department, Faculty of Science, P.O.B. 331,
 Bijeni\v{c}ka c.32,
\\ HR-10002 Zagreb, Croatia \\[3mm] 

\date{\today}
\bigskip

\end{center}
\setcounter{page}{1}
\bigskip



\begin{center}
{\bf Abstract}\\
\end{center}

A multispecies model of Calogero type in $D\geq 1$ dimensions is constructed. The model includes
harmonic, two-body and three-body interactions. Using the underlying conformal $SU(1,1)$ algebra, 
we find the exact eigenenergies corresponding to a class of the exact global collective states.
Analysing corresponding Fock space, we detect the universal critical point at which the model
exhibits singular behaviour.

\bigskip
PACS number(s): 03.65.Fd, 05.30.Pr\\

Keywords: multispecies Calogero model,  SU(1,1) algebra.



\section{Introduction}
The (rational) Calogero model [1] is one of the most famous and exhaustively studied examples of
exactly solvable systems. It describes $N$ identical particles on the line which interact through
an inverse-square two-body interaction and are subjected to a common confining harmonic force.
Starting from the inception more than thirty years ago, the model and its various descedants continue
to be of interest for both physics  and mathematics community, primarly because they are
connected with a number of mathematical and physical problems, ranging from random matrices
and symmetric polynomials [2] to condensed matter systems [3],  black hole physics [4] and 2D strings [5]. 

 The Calogero model is defined by the following Hamiltonian ($i,j = 1,2,...N$):
\be
H=-\frac{\hbar^2}{2m} \sum_{i} \frac{\partial_i^2}{\partial x_i^2} + \frac{m\omega^2}{2}\sum_{i} x_i^2 +
\frac{\hbar^2 \nu (\nu - 1)}{2m} \sum_{i\neq j }\frac{1}{(x_i-x_j)^2},
\ee
where $\nu (\nu - 1)\geq -1/4$ is the dimensionless coupling constant, $m$ is the mass of particles and $\omega$ is 
the strength of a harmonic confinement potential. The ground state of the Hamiltonian (1) is of the well-known, 
highly correlated, Jastrow form
\be
\Psi_0 (x_1,x_2,\cdots x_N)= \prod_{i<j} |x_i -x_j|^{\nu} e^ {-\frac{m\omega}{2\hbar}\sum_{i}  x_i^2},
\ee
with corresponding ground state energy, which depends on $\nu $ explicitly
\be
E_0=\omega \left(\frac{N}{2} + \frac{\nu N (N-1)}{2}\right).
\ee
One can exactly solve this model and find out the complete set of energy eigenvalues either by following the 
traditional approach [1] or by employing its underlying $S_N$ (permutational) algebraic structure [6]. The later 
(operator) approach is considerably simpler than the original one, yields an explicit
expression for the wavefunctions  and emphasizes the interpretation in terms of generalized
statistics. Namely, the inverse-square potential can be 
regarded as a pure statistical interaction  and the model maps to an ideal gas of particles 
obeying fractional Haldane statistics [7], with  coupling constant $\nu$ playing the role of  
Haldane statistical parameter. In Haldane's formulation, however, there is the possibility of having particles of 
different species with a mutual statistical  parameter depending on the species coupled.
This suggests the generalization of the single-species Calogero model to the multispecies Calogero model.
Distinguishability of the species can be introduced by allowing particles to 
have different masses and different couplings to each other. The novel feature of 1D multispecies Calogero model 
is appearance of the long-range three-body interaction. Under certain conditions this three-body interaction can be eliminated
from the Hamiltonian (but only in 1D!). 

Further generalization of Calogero model (1) can be achieved by formulating the model in dimensions higher than one. 
In a case of a single-species model in D dimensions, some exact eigenstates (including the ground state) are known 
but the complete solution of the problem is still lacking. Some progress has been achieved only recently for a class of 2D
models [8]. Usually, the inevitable appearance of the three-body interaction in $D>1$ is the main obstacle which 
makes any analysis of such a model(s) highly nontrivial.

The aim of the present article is to present and discuss some aspects of the model which incorporates both generalizations 
mentioned above. In Sec.2 we define a multispecies Calogero model in D dimensions. We succeeded in finding  a class of, 
but not all, exact eigenstates  of the model  Hamiltonian which corresponded to global collective states.  
The analysis relied heavily on the SU(1,1) algebraic structure of the  Hamiltonian and utilized Fock space representation.  
In Sec.3 we briefly discuss some other interesting features of the model. We rewrite the Hamiltonian in a simple form and 
comment on universal critical point.  Section 4 is concluding section. 



\section{A model Hamiltonian: some exact eigenstates and eigenenergies}         
Following the lines of reasoning already sketched in Introduction, we  define the most general Calogero-type Hamiltonian in  
$D$ dimensions [9],  describing $N$ distinguishable Calogero-like particles which interact  with 
two-body and three-body interactions, as ( we put $\hbar =1$ and $i,j = 1,2,...N$) 
$$
H=-\frac{1}{2}\sum_{i}\frac{1}{m_{i}} \vec{\nabla}_{i}^{2}+ 
\frac{{\omega}^{2}}{2} \sum_{i} m_{i} \vec{r}_{i}^{2}
 + \frac{1}{2} \sum_{i<j} \frac{\nu_{ij}(\nu_{ij}+D-2)}{{|\vec{r}_{i}-\vec{r}_{j}|}^{2}} (\frac{1}{m_{i}} + 
 \frac{1}{m_{j}})+
$$
\be
+ \frac{1}{2} \sum_{i \neq j, i\neq k} \frac{\nu_{ij} \nu_{ik} (\vec{r}_{i}-\vec{r}_{j}) (\vec{r}_{i}-\vec{r}_{k})}{m_{i}  	
{|\vec{r}_{i}-\vec{r}_{j}|}^{2} {|\vec{r}_{i}-\vec{r}_{k}|}^{2}}.
\ee
Here, $m_{i}$ are masses of the particles, $\omega$ is the frequency of the harmonic 
potential and  $\nu_{ij}=\nu_{ji}$ are the statistical parameters between particles $i$ and $j$. \\
It can be shown that the ground state $\Psi_{0}$, obeying $H \Psi_{0} = E_{0} \Psi_{0}$, is of generalized Jastrow form (2)
\be
\Psi_{0}(\vec{r}_{1},...,\vec{r}_{N})=  \prod_{i<j}{|\vec{r}_{i}-\vec{r}_{j}|}^{\nu_{ij}} e^{-\frac{\omega}{2}
 \sum_{i} m_{i} {\vec{r}}_{i}^{2}} \equiv \Delta e^{-\frac{\omega}{2}\sum_{i} m_{i} {\vec{r}}_{i}^{2}}
\ee
and the ground state energy $E_{0}$ generalizes Calogero result (3)
\be
E_{0} = \omega ( \frac{N D}{2} + \sum_{i<j} \nu_{ij})\equiv \omega  \epsilon_0
\ee
Notice that for  $\nu_{ij}=\nu $, $m_i=m$ and $D\neq 1$ Eq.(5) smoothly goes to exact ground state of the 
Calogero-Marchioro model.
For $D=1$, the three-body term in (4) identically vanish if  $\nu_{ij}=\nu$ and
$m_i=m$ (single-species Calogero model [1]) or if there exists a certain relation between masses and coupling constants,
of the form $\nu_{ij}=\alpha m_i m_j $  (multispecies Calogero model [10]). This happens because  identity 
$\sum \frac{1}{(x_i-x_j)(x_i-x_k)}=0$, which holds in 1D only.
Unlike in one dimension, however, it does not vanish in higher dimensions (there is no analogouos identity for $D\neq 1$)
 and plays a crucial role in the analysis 
that is to follow.\\
In order to simplify the analysis, we perform the non-unitary  transformation on $\Psi_{0}$, namely 
$\tilde{\Psi}_{0} = \Delta^{-1} \Psi_{0}$. It generates a similarity transformation which leads to an another
 Hamiltonian $\tilde{H} = \Delta^{-1}H \Delta $. We find $\tilde{H}$ as
$$
\tilde{H} =    \omega^{2}\left ( \frac{1}{2} \sum_{i} m_{i} {\vec{r}}_{i}^{2} \; \right ) -  
\left(\; \frac{1}{2} \sum_{i} \frac{1}{m_{i}} {\vec{\nabla}_{i}}^{2} +   
\sum_{i<j} \nu_{ij}\frac{(\vec{r}_{i}-\vec{r}_{j})}  {{|\vec{r}_{i}-\vec{r}_{j}|}^{2}} 
(\frac{1}{m_{i}} \vec{\nabla}_{i} - \frac{1}{m_{j}} \vec{\nabla}_{j}) \;\right ) 
$$
\be
 \equiv   {\omega}^{2} T_{+} - T_{-}.
\ee
In addition to $T_{\pm}$,we also introduce dilatation operator 
$T_0=\frac{1}{2} (\sum_{i}\vec{r}_{i} \vec{\nabla}_{i} + {\varepsilon}_{0})$.
The operators $T_{\pm} ,\; T_{0}$ satisfy the $ SU(1,1) $ algebra.\\
It is convenient to introduce the center-of-mass coordinate $ \vec{R} $ and 
the relative coordinates $ \vec{\rho}_{i} $:
$$
\vec{R} = \frac{1}{M}  \sum_{i} m_{i} \vec{r}_{i}, \qquad \vec{\nabla}_{R} = \sum_{i}  \vec{\nabla}_{i},
$$
\be
\vec{\rho}_{i} = \vec{r}_{i} - \vec{R}, \qquad \vec{\nabla}_{\rho_i}=
\vec{\nabla}_{i}-\frac{m_i}{M}\vec{\nabla}_{R}.
\ee
Using Eqs.(7) and (8) we define creation (+) and annihilation (-) operators 
$$
\vec{A_{1}}^{\pm} = \frac{1}{\sqrt{2}} (\sqrt{M \omega} \vec{R} \mp \frac{1}{\sqrt{M \omega}} \vec{\nabla}_{R}),
$$
\be
{A_{2}}^{\pm} = \frac{1}{2} ( \frac{T_{-}}{\omega} + \omega T_{+}) \mp T_{0},
\ee
which satisfy the following commutation relations ($\alpha, \beta =1,2,\cdots D$) :
$$
[A_{1,\alpha}^{-}, A_{1,\beta}^{+}] = \delta_{\alpha\beta},\qquad 
 [A_{1,\alpha}^{-}, A_{1,\beta}^{-}] = 
[A_{1,\alpha}^{+}, A_{1,\beta}^{+}] = 0, 
$$
$$
[\vec{A_{1}}^{-}, {A_{2}}^{+}] = \vec{A_{1}}^{+}, \qquad
[A_{2}^{-},\vec{A_{1}}^{+}] = \vec{A_{1}}^{-},
$$
$$
[A_{2}^{-}, {A_{2}}^{+}] = \frac{\tilde{H}}{\omega}, \qquad 
[\tilde{H}, \vec{A_{1}}^{\pm}] = \pm \omega \vec{A_{1}}^{\pm} ,\qquad 
$$
\be
[\tilde{H}, {A_{2}}^{\pm}] = \pm 2 \omega{A_{2}}^{\pm}.
\ee
They act on the Fock vacuum $|\tilde{0} \rangle \propto \tilde{\Psi}_{0}(\vec{r}_{1},...,\vec{r}_{N})$ as
\be
\vec{A_{1}}^{-}|\tilde{0} \rangle = A_{2}^{-}|\tilde{0} \rangle = 0 , \qquad  \langle\tilde{0} |\tilde{0} \rangle = 1.
\ee
The excited states in the Fock space, corresponding to global collective states, are of the form  
\be
{\left(A_{1,1}^{+}\right)}^{n_{1,1}}\cdots (A_{1,D}^{+})^{n_{1,D}} {\left(A_{2}^{+}\right)}^{n_{2}}
|\tilde{0} \rangle \equiv
\prod_{\alpha=1}^{D}(A_{1,\alpha}^{+})^{n_{1,\alpha}} {\left(A_{2}^{+}\right)}^{n_{2}}|\tilde{0} \rangle,
\ee
where $n_{1,\alpha}=0,1,2... (\forall \alpha)$ and $n_2=0,1,2...$\\
The repeated action of the operators ${A^{+}_{1,\alpha}}$ 
on the vacuum $|\tilde{0} \rangle $ reproduces, in the coordinate representation,
Hermite polynomials $H_{n_{1,\alpha}}(R_{\alpha}{\sqrt {M\omega}})$. Similarly, the repeated  
action of the operator $ A_{2}^{+}$ on the vacuum $|\tilde{0} \rangle $ reproduces  hypergeometric function, which reduces to 
associated Laguerre polynomials 
$  L_{n_2 + \varepsilon_{0}-1}^{\varepsilon_{0}-1} (2 \omega T_{+}$) for certain values of parameters.\\
The states (12) are eigenstates of the $\tilde{H}$ with the energy eigenvalues (cf. last two equations in Eqs.(10))
\be
E_{n_{1,\alpha}; n_{2}} = \omega \left( \sum_{\alpha = 1}^{D} n_{1, \alpha} + 2n_{2} + {\varepsilon}_{0}\right).
\ee
This is the part of the complete spectrum  which corresponds to center-of-mass states and global dilatation states,
respectively. We note in passing that this states, Eq.(12), are perfectly normalizable 
 (i.e. quadratically integrable) and physically acceptable for  both
Hamiltonians $\tilde{H}$ and $H$, provided that $\epsilon_0 >\frac{D}{2}$.



\section{Miscellaneous remarks }
It is evident from Eqs.(10) that the modes described by $\vec{A_{1}}^{\pm}$ and $A_{2}^{\pm}$ are still coupled. By introducing
another set of of the creation and annihilation operators $\{B_{2}^{+},B_{2}^{-}\}$
\be
B_{2}^{\pm} = {A_{2}}^{\pm} - \frac{1}{2}(\vec{A_{1}}^{\pm})^{2} ,
\ee
one can show that the center-of-mass motion decouples completely
\be
[{A_{1, \alpha}}^{\pm}, {B_{2}}^{\mp}] = 0 .
\ee
The Hamiltonian $\tilde{H}$  separates as $\tilde{H}=\tilde{H}_{CM} + \tilde{H}_{R}$, with  
$$
\tilde{H}_{R}= \omega [B_{2}^{-},B_{2}^{+}] ,\qquad
 [\tilde{H}_{R},B_{2}^{\pm}] = \pm 2 \omega B_{2}^{\pm},
$$
\be
\tilde{H}_{CM}=\frac{1}{2}\omega \sum_{\alpha = 1}^{D} \{ A_{1, \alpha}^{-} , 
{A_{1, \alpha}}^{+} \}_+ , \qquad
[\tilde{H}_{CM}, \vec{A_{1}}^{\pm}] = \pm \omega \vec{A_{1}}^{\pm}.
\ee
The corresponding Fock space (12) now splits  into the CM-Fock space, spanned by 
$\prod_{\alpha}(A_{1,\alpha}^{+})^{n_{1,\alpha}} |\tilde{0} \rangle_{CM}$ and the R-Fock space, spanned by
$B_{2}^{+ n_2}|\tilde{0}\rangle_{R} $. The respective vacua are 
$|\tilde{0}\rangle_{CM}\propto  e^{-\frac{\omega}{2} M {\vec{R}}^{2}} $ and 
$ |\tilde{0}\rangle_{R} \propto e^{-\frac{\omega}{2} 
\sum_i m_{i} \vec{\rho}_{i}^{2}} $. \\

Closer inspection of the R-Fock space of the Hamiltonian $\tilde{H}_{R}$ reveals the existence of the 
universal critical point defined either by  the  null-vector 
\be
\frac{_R\langle \tilde{0} | B_{2}^{-} B_{2}^{+}| \tilde{0} \rangle_R}{_R\langle \tilde{0} |\tilde{0} \rangle_R} = 0 
\Rightarrow \frac{(N-1) D}{2} + \frac{1}{2}\sum_{i\neq j } \nu_{ij}  = 0,
\ee
or, equivalently, by the zero-energy condition
\be
E_{0R}= \frac{(N-1)D}{2} + \frac{1}{2}\sum_{i\neq j } \nu_{ij}  = 0.
\ee
At the critical point  the system described by  $\tilde{H}_{R}$  collapses completely. 
This means that the relative coordinates, the relative momenta and the relative energy  are all zero at this 
critical point. There survives only one oscillator, describing the motion of the
centre-of-mass. This critical point  was first noticed by analysing  Gram matrices of the scalar products of Fock-space states  
in ordinary Calogero model [11] and further confirmed by large-N collective field 
theory approach to the same model  in Ref.[12].\\ 
For the initial Hamiltonian (4),  which is not unitary  equivalent to  $\tilde{H}$, the conditions (17,18) demand
that some $\nu_{ij} < 0$ and, consequently,  the norm of the wave function (5) blows up at the critical point. 
For $\nu_{ij}$ negative but greater than the critical values (17,18), the wave function is singular 
at coincidence points but still quadratically integrable.
 



\section{Conclusion}
In summary, we have studied a non-trivial many-body Hamiltonian  of Calogero type in D dimensions (4), with two- and 
three-body interactions among non-identical particles. By applying the similarity transformation, we have obtained the simpler
Hamiltonian $\tilde{H}$ (7), on which we have performed the Fock-space analysis and found some of its excited (collective) 
states (12) and their energies (13). The spectrum of collective modes is linear, equidistant and degenerate. 
By splitting the Fock space into the CM-Fock space and the R-Fock space (16)
we have detected  the universal critical point (17,18) at which the system exhibits singular behaviour.\\
Unfortunately, within our algebraic treatment, we are unable to construct the rest of the eigenstates of the Hamiltonian (7) 
which parallel those in Calogero model (1). Therefore, the question of the full solvability of the model Hamiltonian (4) still remains open. 
Nevertheless, we hope that our analysis sheeds some light on the  such kind of higher-dimensional models, that is the models 
with similar underlying conformal SU(1,1) symmetry. \\

{\bf Acknowledgments}\\

One of the authors (M.M.) would like to thank Prof. \v C. Burdik for kind invitation and hospitality during the XIII 
International Colloquium " Integrable Systems and Quantum Groups", Prague, June 17-19, 2004.\\
This work was supported by the Ministry of Science and Technology of the Republic of Croatia under 
contracts No. 0098003 and No. 0119261.

\end {document}